# An Extended *T-A* Formulation Based on Potential-Chain Recursion for Electromagnetic Modeling of Parallel-Wound No-Insulation HTS Coils

Zhe Pan[1], Qi Xu[1], Ruixiang Wang[1], Zhenghao Jin[1] and Jianzhao Geng[1,*]

[1] Wuhan National High Magnetic Field Center, Huazhong University of Science and Technology, Wuhan, People's Republic of China

E-mail: gengjianzhao@hust.edu.cn

## Abstract

Parallel-wound no-insulation (PW-NI) high-temperature superconducting (HTS) coils significantly reduce charging delay while maintaining excellent self-protection capability, demonstrating great potential for high-field applications. Existing models that couple the *T-A* formulation with equivalent circuits have demonstrated high accuracy in electromagnetic analysis of PW-NI coils. However, eliminating the computational overhead caused by frequent variable mapping and data exchange between electromagnetic and circuit modules is important for improving computational efficiency, particularly in long-duration transient simulations of large-scale magnets. To address this issue, an extended *T-A* formulation based on potential-chain recursion, termed PCR-TA, is proposed. By directly embedding inter-tape current sharing and radial current bypass behaviors into the finite-element framework, this method computes the transient electromagnetic response of PW-NI coils without requiring an explicit equivalent circuit model. Building upon it, a multi-scale approach is further developed for large-scale PW-NI coils. The validity of the proposed method and its multi-scale extension is verified through comparisons with experimental measurements and field-circuit coupled modeling results. Comparative analyses demonstrate that the PCR-TA method achieves a speedup of approximately 2.4 over the field-circuit coupled method, whereas its multi-scale extension further increases this speedup to roughly 5.8. Furthermore, the PCR-TA method is extended to model the continuous transition of PW-NI coils from power-supply charging to closed-loop operation. The results reveal current imbalances with spatially varying characteristics during the free decay. Under an external radial AC magnetic field, asymmetric transient dynamic resistance arises in parallel tapes, driving current redistribution. This work provides an efficient method and tool for the electromagnetic modeling of PW-NI coils under both driven and closed-loop operating conditions.

## 1. Introduction

In recent years, no-insulation (NI) high-temperature superconducting (HTS) coils have attracted considerable attention in the development of high-field magnets due to their excellent self-protection capabilities and high engineering current density [1-6]. The elimination of inter-turn insulation allows current to bypass local normal zones, thereby significantly enhancing the magnet's thermal stability [7-9]. However, inter-turn radial currents induce substantial charging delays and additional losses [10-14]. To mitigate this issue, the parallel-wound no-insulation (PW-NI) structure has been proposed [15]. By connecting multiple superconducting tapes in parallel, this design reduces the equivalent coil inductance, thereby shortening charging delays and alleviating the stringent length requirements for individual tapes [16-19]. To date, PW-NI technology has been successfully implemented in several high-field magnet projects. Notable examples include the 35.6 T all-superconducting user magnet developed by the Institute of Electrical Engineering, Chinese Academy of Sciences [20]; the 20 T toroidal field (TF) model coil fabricated by MIT and Commonwealth Fusion Systems (CFS) [21, 22]; and the 24.4 T six-tape PW-NI magnet constructed by Tokamak Energy [23].

Due to inductance differences among parallel tapes, PW-NI coils inevitably exhibit significant transient current non-uniformity, which has attracted considerable research attention. Kobayashi *et al.* extended the partial element equivalent circuit (PEEC) model to PW-NI coils to investigate the impact of current imbalance on thermal stability and central magnetic field [24]. Lee *et al.* employed a multi-tape turn distributed circuit model to study the effects of tape properties and contact resistance on the critical current of PW-NI coils [25]. Field-based formulations, such as the *H*-formulation [26-29] and the *T-A* formulation [30-32], are well-suited for capturing screening current effects and local current density distributions because they directly solve Maxwell's equations. However, in PW-NI coils, the complex inter-tape current sharing and radial current paths prevent these formulations from directly determining the transient transport current in each parallel tape. To address this issue, Fu *et al.* coupled the PEEC model with the *T-A* formulation to simulate transient electromagnetic responses during fast charging/discharging [33] and quench processes [34]. Based on the *T-A* formulation, Liu *et al.* proposed an improved field-circuit coupled model that avoids the calculation of the inductance matrix and reveals the dominant role of terminal resistance in steady-state current non-uniformity [35]. Similarly, Chen *et al*. integrated an equivalent circuit model into the *T-A* formulation, thereby bypassing the inductance extraction process and investigating the coupling losses and hoop stresses in large-scale magnets [36]. Recently, Chen *et al.* further introduced the Coulomb gauge and employed reduced-order nodal current conservation equations to extend the *T-A* formulation to a three-dimensional helical framework, which was combined with external circuits to simulate a 3D dual-tape PW-NI coil [37].

To date, *T-A* formulation coupled with explicit circuit models (e.g., PEEC or equivalent circuits) has demonstrated exceptional validity and fidelity in capturing both macroscopic current redistribution and local electromagnetic properties in PW-NI coils [33, 35, 38, 39]. To further enhance computational efficiency, addressing the overhead arising from repetitive variable mapping and data exchange between the electromagnetic and circuit modules remains a critical objective, particularly for long-duration transient simulations of large-scale magnets.

Therefore, this paper proposes an efficient electromagnetic modeling approach for PW-NI coils based on an extended *T-A* formulation derived through potential-chain recursion (PCR-TA). Instead of constructing an equivalent circuit model, the PCR-TA method integrates the inter-tape current sharing and radial current bypass behaviors within a unified finite-element framework, enabling direct calculation of transient electromagnetic responses. Building upon this, a multi-scale model is further developed to significantly enhance the computational efficiency for large-scale problems without compromising accuracy. Furthermore, the PCR-TA method is extended to enable the continuous simulation of PW-NI coils under both driven and closed-loop operating conditions. This paper is organized as follows. The development of the numerical modeling framework, including the PCR-TA method and its multi-scale approach, is detailed in Section 2. The accuracy and efficiency of the proposed method and the multi-scale approach are validated by comparing them with experimental measurements and field-circuit coupled modeling results in Section 3. The transient current evolution of closed-loop PW-NI coils under both free-decay conditions and external radial AC magnetic field disturbances is investigated in Section 4. The main conclusions of this work are summarized in Section 5.

## 2. Numerical modeling method

### 2.1 T-A formulation

The *T-A* formulation describes electromagnetic fields by coupling the magnetic vector potential $A$ and the current vector potential $T$. The governing equations of the *T-A* formulation are given as follows [30, 31]:

$$\begin{cases} \nabla^2 A = -\mu J \\ \nabla \times \rho_{\text{HTS}} \nabla \times T = -\dfrac{\partial B}{\partial t} \end{cases} \quad (1)$$

where $\mu$ and $\rho_{\text{HTS}}$ are the magnetic permeability and the resistivity of the superconducting layer, respectively. Based on the thin-film approximation [40], the superconducting layer is assumed to have negligible thickness. Consequently, the current vector potential $T$ has only a component $T_r$ perpendicular to the tape surface. The current density $J$ is assumed to be uniform across the thickness and has only the azimuthal component $J_\varphi$. Under these conditions, for a 2D axisymmetric model, the relationship between $J_\varphi$ and $T_r$ is expressed as:

$$J_\varphi = \frac{\partial T_r}{\partial z} \quad (2)$$

By substituting equation (2) into the radial component of Faraday's law, the scalar governing equation for the *T*-formulation within the superconducting tape is derived as:

$$\frac{\partial E_\varphi(J_\varphi)}{\partial z} = \frac{\partial B_r}{\partial t} \quad (3)$$

where the nonlinear relationship between the electric field and the current density follows the *E-J* power law:

$$E_\varphi = \rho_{\text{HTS}} J_\varphi = \frac{E_0}{J_c(B)} \left| \frac{J_\varphi}{J_c(B)} \right|^{n-1} J_\varphi \quad (4)$$

where the characteristic electric field $E_0$ is $1 \times 10^{-4}$ V·m$^{-1}$, and $J_c(B)$ denotes the critical current density, as defined by the modified Kim model. This model accounts for the anisotropic dependence of the critical current density on the magnetic flux density, expressed as follows [41]:

$$J_c(B) = \frac{J_{c0}}{\left(1 + \sqrt{(mB_\parallel)^2 + B_\perp^2}/B_0\right)^\alpha} \quad (5)$$

where $B_\parallel$ and $B_\perp$ represent the magnetic flux density components parallel and perpendicular to the superconducting tape, respectively. The terms $m$, $\alpha$, and $B_0$ are material-dependent parameters, with $m = 0.0605$, $\alpha = 0.7580$, and $B_0 = 0.103$ T [42].

To solve the partial differential equation (3), a weak formulation is established within the finite element framework. By multiplying both sides of the equation (3) by a test function $\tilde{T}$ and integrating over the superconducting layer, the following expression is obtained [43]:

$$\int_a^b \left( E_\varphi \frac{\partial \tilde{T}}{\partial z} + \frac{\partial B_r}{\partial t} \tilde{T} \right) dz = \tilde{T} \cdot \left[ E_\varphi \right]_{z=a}^{z=b} \quad (6)$$

where $a$ and $b$ denote the $z$-coordinates of the two edges along the width of the superconducting tape,

respectively. The right-hand term characterizes the energy conversion process within the *T*-formulation, indicating that the boundary electric field directly governs the internal electromagnetic field distribution of the superconducting tape. In conventional *T-A* formulation, the scalar potential contribution from the external power supply is typically neglected, and this boundary electric field is defined via the Neumann boundary condition as $E_\varphi = -\partial A_\varphi/\partial t$. However, this treatment proves inaccurate for current-sharing problems, necessitating a modification to this boundary condition as follows [44]:

$$E_\varphi = -\frac{\partial A_\varphi}{\partial t} - \frac{U}{2\pi r} \tag{7}$$

where $U$ is the terminal voltage across the superconducting tape, and $r$ represents its radius. The modification term $-U/(2\pi r)$ represents the electrostatic potential gradient generated by the external power supply. Due to the introduction of the new unknown variable $U$, a corresponding global equation is required to constrain the transport current $I_{op}$:

$$I_{op} = \iint_S J_\varphi \, dS = d_{tape} \cdot \int_a^b J_\varphi \, dz \tag{8}$$

where $I_{op}$ denotes the transport current provided by the power supply, and $d_{tape}$ represents the thickness of the superconducting tape.

*2.2 Potential-chain recursion*

In this section, potential-chain recursion is proposed to extend the *T-A* formulation, enabling the description of current redistribution and radial bypass behavior in PW-NI coils within a unified finite-element framework. Notably, the PCR-TA method can be applied to pancake coils with various winding configurations.

Figure 1 illustrates the current paths in a 3-turn dual-tape PW-NI coil as an example. Here, Tape A is the inner tape at each turn, while Tape B is the outer tape. $I_{in1}$ and $I_{in2}$ denote the input currents of Tape A and Tape B, respectively, whereas $I_{out1}$ and $I_{out2}$ denote their corresponding output currents. $I_{a,i}$ and $I_{b,i}$ represent the azimuthal currents of Tape A and Tape B in the *i*-th turn. $I_{tt,i}$ and $I_{turn,i}$ denote the intra-turn and inter-turn radial currents of the *i*-th turn, respectively. In addition, $U_{a,1}$ and $U_{b,1}$ denote the terminal voltages of Tape A and Tape B in the first turn, respectively (the remaining turns are omitted for clarity).

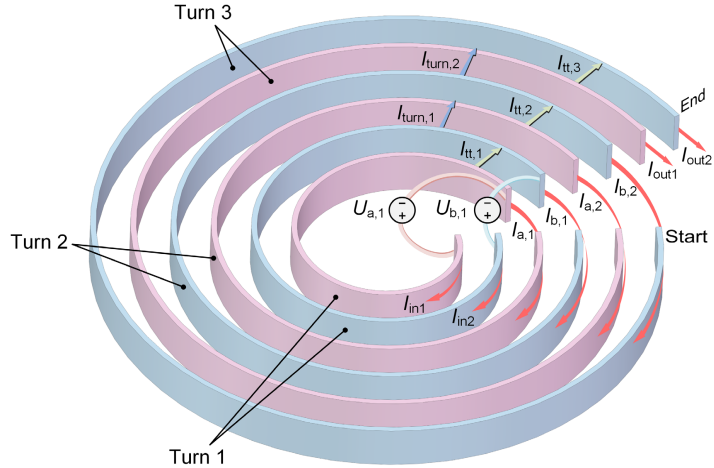

**Figure 1.** Geometry and current paths of a dual-tape PW-NI coil. The figure defines the inner tape (Tape A) and outer tape (Tape B) in each turn, illustrating the input and output currents, azimuthal currents ($I_{a,i}$ and $I_{b,i}$), intra-turn ($I_{tt,i}$) and inter-turn ($I_{turn,i}$) radial currents, and terminal voltages ($U_{a,i}$ and $U_{b,i}$).

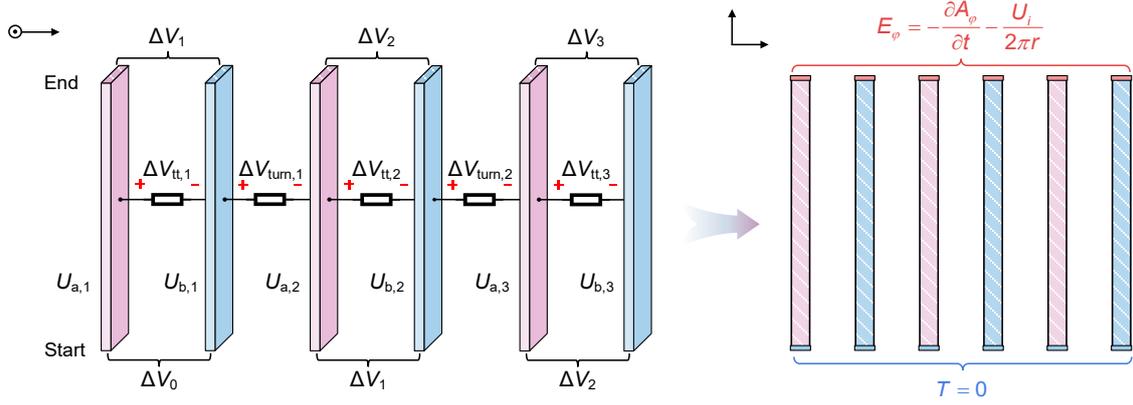

**Figure 2.** Schematic of the PCR-TA method. The diagram illustrates the inter-tape potential differences ($\Delta V_i$) used to quantify current sharing and radial currents, as well as the resulting driving voltages ($V_{tt,i}$ and $V_{turn,i}$) for the intra-turn and inter-turn radial currents within the finite-element framework.

To accurately quantify the current sharing and radial leakage currents between the parallel tapes, $\Delta V_i$ is defined as the potential difference between the tapes at the end of the $i$-th turn, as shown in figure 2. Owing to the geometric continuity of the helical winding, the potential difference between the tapes at the beginning of the $i$-th turn is equal to that at the end of the preceding turn. Considering the cumulative voltage drop along the helical path, the recursive relation for $\Delta V_i$ can be expressed as:

$$\Delta V_i = \Delta V_{i-1} - \left(U_{a,i} - U_{b,i}\right) \quad (9)$$

The potential difference between the parallel tapes at the inlet is given by:

$$\Delta V_0 = I_{in2} \cdot R_{in2} - I_{in1} \cdot R_{in1} \quad (10)$$

where $R_{in1}$ and $R_{in2}$ represent the input joint resistances of Tape A and Tape B, respectively.

Based on the defined inter-tape potential difference, the driving voltages for the intra-turn and inter-turn radial currents can be further determined. The driving voltage for the intra-turn radial current flowing from Tape A to Tape B within the same turn is given by:

$$\Delta V_{tt,i} = \frac{1}{2}\left(\Delta V_i + \Delta V_{i-1}\right) \quad (11)$$

Meanwhile, the driving voltage for the inter-turn radial current flowing from Tape B of the $i$-th turn to Tape A of the $(i+1)$-th turn is given by:

$$\Delta V_{turn,i} = -\Delta V_i + \frac{1}{2}\left(U_{a,i+1} + U_{b,i}\right) \quad (12)$$

Consequently, the intra-turn and inter-turn radial currents can be readily calculated based on these driving voltages:

$$\begin{cases} I_{tt,i} = \dfrac{\Delta V_{tt,i}}{\rho_{ct}/(2\pi r_{2i-1})} \\ I_{turn,i} = \dfrac{\Delta V_{turn,i}}{\rho_{ct}/(2\pi r_{2i})} \end{cases} \quad (13)$$

where $\rho_{ct}$ is the equivalent contact resistivity of the coil; $r_{2i-1}$ and $r_{2i}$ are the radii of Tape A and Tape B in the $i$-th turn, respectively; and $w$ is the tape width.

For an $N$-turn dual-tape PW-NI coil, the model inherently incorporates $2N+1$ global scalar variables to characterize the circuit states, including the input current of Tape A ($I_{in1}$) and the voltage variables along the spiral path for both tapes, $U_{a,i}$ and $U_{b,i}$ ($i = 1, 2, ..., N$). The azimuthal and radial current distributions bridge the physical coupling between these global circuit variables and the electromagnetic field. Specifically, this coupling is realized by enforcing current continuity between the azimuthal currents computed using the *T-A* formulation and the intra-turn and turn-to-turn radial currents driven by the terminal voltage variables. To determine these variables, a global system of constraint equations is formulated by enforcing current continuity at characteristic locations of the winding, as follows:

1) Input terminal constraints ($i = 1$). At the input side, the operating current $I_{op}$ is partitioned between the two parallel tapes. This distribution inherently satisfies the initial equilibrium among the azimuthal currents and the intra-turn and inter-turn radial currents:

$$\begin{cases} I_{in1} - I_{a,1} - I_{tt,1} = 0 \\ (I_{op} - I_{in1}) - I_{b,1} + I_{tt,1} - I_{turn,1} = 0 \end{cases} \quad (14)$$

2) Chained recursion for intermediate turns ($i = 2, 3, ..., N-1$). For the intermediate turns, the current continuity dictates a dynamic balance between the azimuthal and radial currents. In accordance with current conservation, the total current exiting the (i-1)-th turn must strictly equal that entering the i-th turn:

$$\begin{cases} (I_{a,i-1} + I_{turn,i-1}) - I_{a,i} - I_{tt,i} = 0 \\ (I_{b,i-1} + I_{tt,i}) - I_{b,i} - I_{turn,1} = 0 \end{cases} \quad (15)$$

3) Output terminal constraints ($i = N$). For the final turn, the inter-turn radial current path is terminated. The potential recursion sequence is closed by applying a voltage constraint governed by the output joint resistances:

$$\begin{cases} (I_{a,N-1} + I_{turn,i-1}) - I_{a,N} - I_{tt,N} = 0 \\ \Delta V_N - (I_{a,N} \cdot R_{out1} - I_{b,N} \cdot R_{out2}) = 0 \end{cases} \quad (16)$$

where $R_{out1}$ and $R_{out2}$ denote the output joint resistances of Tape A and Tape B, respectively. The azimuthal currents $I_{a,i}$ and $I_{b,i}$ are defined as the integrals of the azimuthal current density $J_\varphi$, obtained from the *T-A* formulation, over the corresponding superconducting domains:

$$I_{a,i} = d_{tape} \cdot \int J_\varphi \, dl \quad (17)$$

The above set of equations is strongly coupled with the *T-A* formulation and solved within the same stiffness matrix. As a result, the electromagnetic behavior of the PW-NI coil can be resolved entirely within the finite-element framework without introducing additional equivalent circuit models or cross-module data exchange. This approach significantly improves computational efficiency and, importantly, prevents the non-physical instabilities that often arise in segregated or loosely coupled simulation schemes.

*2.3 Multi-scale modeling*

In this section, a multi-scale extension of the PCR-TA method is introduced for the efficient electromagnetic simulation of large-scale PW-NI magnets. This approach is employed to accelerate large-scale computations within the *T-A* formulation [31]. The multi-scale approach divides all turns of the PW-NI coil into analyzed and non-analyzed turns. Each tape in the analyzed turns has its terminal voltage individually defined, and the azimuthal current distribution is solved using the *T-A* formulation. The terminal voltage and azimuthal current of the non-analyzed turns are obtained via linear interpolation between adjacent analyzed turns. For Tape A, a non-analyzed turn *j* located between two adjacent analyzed turns *i* and *k* is considered. The terminal voltage and azimuthal current of turn *j* can then be expressed as:

$$\begin{cases} U_{a,j} = U_{a,i} \cdot \omega_i(r_j) + U_{a,k} \cdot \omega_k(r_j) \\ I_{a,j} = I_{a,i} \cdot \omega_i(r_j) + I_{a,k} \cdot \omega_k(r_j) \end{cases} \quad (18)$$

where $\omega_i(r_j)$ and $\omega_k(r_j)$ are the linear interpolation weighting functions based on the radial position, defined as:

$$\omega_i(r_j) = \frac{r_k - r_j}{r_k - r_i}, \quad \omega_k(r_j) = \frac{r_j - r_i}{r_k - r_i} \quad (19)$$

Different boundary conditions are applied to the analyzed and non-analyzed turns. For the analyzed turns, Neumann boundary conditions are directly imposed. For non-analyzed turns, Dirichlet boundary conditions are enforced in a weak form via a penalty function, with the corresponding variational virtual work term $\delta W_p$ given as:

$$\delta W_p = \int_l \beta \cdot (T - I_{a,j} / d_{tape}) \cdot 2\pi r \tilde{T} \, dl \quad (20)$$

where $\beta$ is the penalty parameter that controls the constraint strength, allowing slight numerical relaxation in the interpolated turns to maintain continuity with neighboring analyzed turns while preserving overall solution stability. Figure 3 illustrates the multi-scale approach.

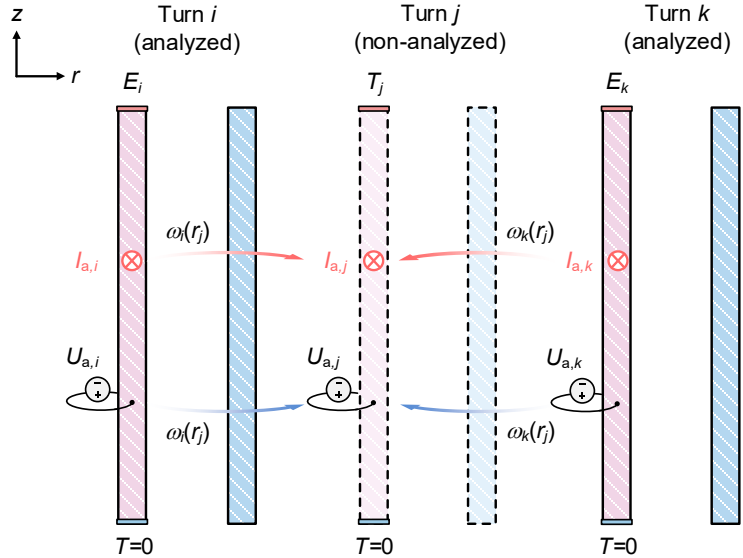

**Figure 3.** Schematic of the multi-scale approach and boundary constraint principle. The azimuthal current and terminal voltage of Tape A in non-analyzed turn *j* are computed from analyzed turns *i* and *k*. Analyzed turns employ Neumann boundary conditions, while non-analyzed turns use Dirichlet boundary conditions.

## 3. Model verification

### 3.1 PCR-TA model

To comprehensively verify the accuracy and applicability of the PCR-TA method, numerical models were developed for validation studies on both dual-tape and three-tape PW-NI coils. The non-uniform current distribution of the dual-tape coil during charging was first simulated and validated against the experimental results reported by Liu *et al.* [35]. In their work, the currents of each parallel tape were measured using Rogowski coils, providing reliable reference data for this study. To accurately reproduce the experimental setup, the simulation model adopts the geometric dimensions and material constitutive parameters consistent with the reference. The specific simulation parameters of the coils are listed in table 1. All numerical simulations in this work were performed using COMSOL Multiphysics on a PC equipped with an Intel(R) Core(TM) i9-14900HX CPU @ 2.20 GHz, 32 GB RAM, running Windows 10 (64-bit).

**Table 1.** Specifications of the 30-turn dual-tape PW-NI coil reported by Liu *et al.* [35]

| Parameters | Value |
|---|---|
| Number of parallel tapes | 2 |
| Number of turns | 30 |
| Inner diameter | 100 mm |
| Outer diameter | 111.4 mm |
| $I_c$ of coil@77 K | 169 A |
| Width/thickness of the tape | 4/0.095 mm |
| Inductance | 193 µH |
| Contact resistance | 184 µΩ |
| Input joint resistance $R_{in1}/R_{in2}$ | 250/237 nΩ |
| Output joint resistance $R_{out1}/R_{out2}$ | 584/220 nΩ |

Figure 4 illustrates the simulated current distribution in the parallel tapes at the input and output terminals under a ramp rate of 6.92 A/s. As depicted, a significant non-uniformity in the current distribution is observed, primarily due to differences in tape inductance. The PCR-TA model accurately captures this transient current-sharing behavior. Upon the completion of charging, the current reaches a steady state governed by the joint resistance ratios. Furthermore, both the simulated transient behavior and the steady-state results are in good agreement with the experimental measurements reported in Ref. [35].

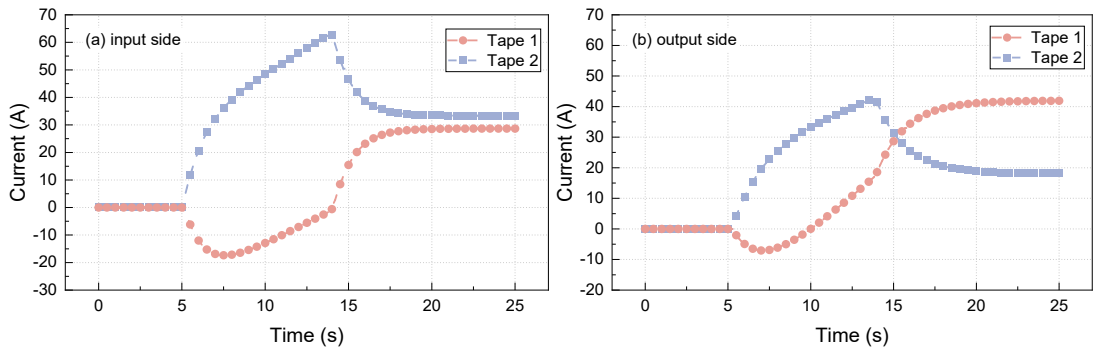

**Figure 4.** Simulated current distribution during the charging process of a dual-tape PW-NI coil: (a) at the input terminal and (b) at the output terminal, with a ramp rate of 6.92 A/s. The experimental data for comparison can be found in Ref. [35].

Figure 5 shows the simulated terminal voltage of the coil during charging. At a ramp rate of 6.92 A/s, the simulated voltage rises rapidly and stabilizes at approximately 1.35 mV. Once the current ramping concludes, the inductive voltage drop vanishes, and the terminal voltage decays exponentially to zero. Furthermore, the simulation results show good agreement with the experimental measurements reported in Ref. [35] for both the voltage amplitude and the dynamic response, validating the accuracy of the PCR-TA model in predicting electromagnetic behavior.

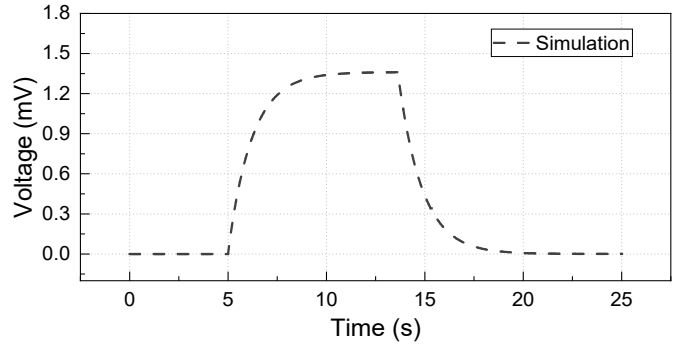

**Figure 5.** Simulated evolution of the coil terminal voltage over time at a charging ramp rate of 6.92 A/s. The experimental data for comparison can be found in Ref. [35].

To demonstrate the scalability and accuracy of the PCR-TA method for multi-tape assemblies, the validation was further extended to a three-tape PW-NI coil. An experimental prototype was fabricated and tested to provide reference data for current distribution. The coil was wound from 4 mm-wide REBCO tapes provided by Shanghai Superconductor Technology Co., Ltd., and charging experiments were conducted in a 77 K liquid nitrogen bath. To accurately monitor the current distribution, the parallel tapes were separated at both terminals, and the individual currents were measured using open-loop Hall current sensors, as shown in figure 6. Furthermore, the equivalent contact resistance of the coil was determined to be 859 μΩ·using the sudden-discharge method [45], and the detailed specifications are summarized in table 2. In addition, we developed a field-circuit coupled model incorporating these parameters following the methodology in Refs. [35, 46], enabling a comparison with the PCR-TA model.

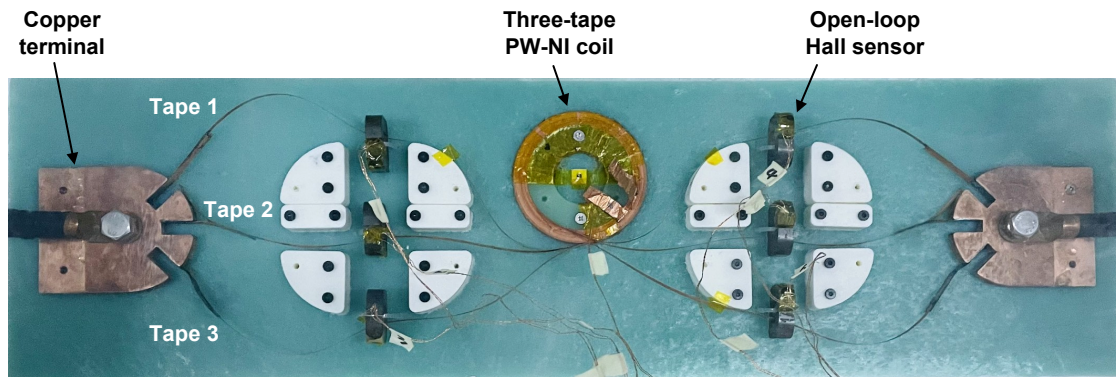

**Figure 6.** Three-tape PW-NI coil prototype and the open-loop Hall sensor test setup for measuring input and output terminal currents.

**Table 2.** Specifications of the three-tape PW-NI coil for experimental validation

| Parameters | Value |
|---|---|
| Number of parallel tapes | 3 |
| Number of turns | 32 |
| Inner diameter | 70 mm |
| Outer diameter | 88.2 mm |
| $I_c$ of coil@77 K | 258 A |
| Width/thickness of the tape | 4/0.095 mm |
| Inductance | 136 μH |
| Input joint resistance | 5.67/5.39/5.31 μΩ |
| Output joint resistance | 5.04/4.93/5.22 μΩ |

Figure 7 compares the current distributions at the input and output terminals obtained from experimental measurements, the field-circuit coupled model, and the PCR-TA model. At a charging rate of 29.16 A/s, the three-tape configuration also exhibits transient current imbalance, accompanied by pronounced local reverse currents. As shown in the figure 7, the transient and steady-state responses calculated by the PCR-TA model are in good agreement with those obtained from the field-circuit coupled model. However, as the current increases, both numerical results deviate from the measured current of Tape 3. This discrepancy may be associated with electromagnetic forces at high currents, which can alter inter-turn contact pressure and lead to non-uniform contact resistivity. Overall, the PCR-TA model shows good agreement with the experimental results and the field-circuit coupled model, validating its accuracy in predicting the electromagnetic behavior of multi-tape PW-NI coils.

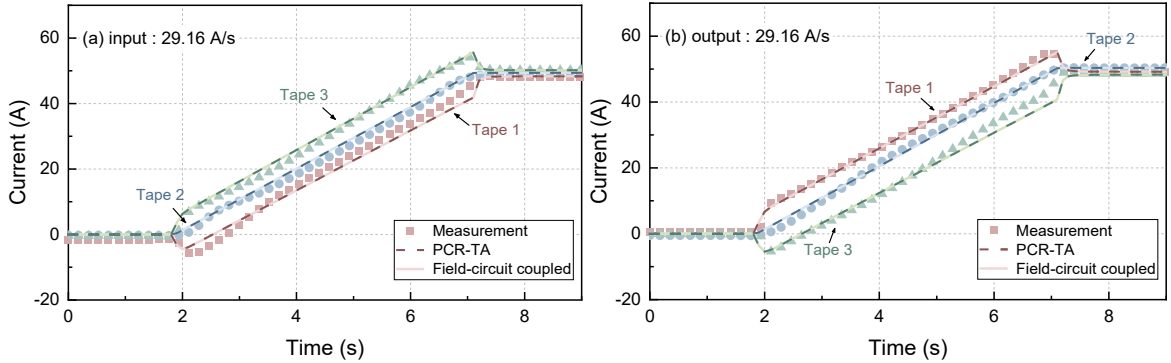

**Figure 7.** Comparison of terminal currents from measurements, the field-circuit coupled model, and the PCR-TA model for a three-tape PW-NI coil at a charging ramp rate of 29.16 A/s: (a) input terminal, (b) output terminal.

*3.2 Multi-scale PCR-TA model*

To verify the accuracy and computational efficiency of the multi-scale approach, a multi-scale PCR-TA model was developed, and its results were benchmarked against those of the full-scale PCR-TA model and the field-circuit coupled model. Specifically, a simulation case for a dual-tape PW-NI coil with 150 turns was constructed, with coil parameters listed in table 3.

**Table 3.** Specifications of the 150-turn dual-tape PW-NI coil

| Parameters | Value |
|---|---|
| Number of parallel tapes | 2 |
| Number of turns | 150 |
| Inner diameter | 60 mm |
| Outer diameter | 116.8 mm |
| $I_c$ of tape@77 K | 233 A |
| Width/thickness of the tape | 4/0.095 mm |
| Contact resistivity $\rho_{ct}$ | 50 μΩ·cm$^2$ |
| Input joint resistance $R_{in1}/R_{in2}$ | 300/300 nΩ |
| Output joint resistance $R_{out1}/R_{out2}$ | 300/300 nΩ |

The selection of analyzed turns and the mesh distribution in the multi-scale PCR-TA model are shown in figure 8. Given that the inner and outer boundary regions of the coil typically exhibit severe radial magnetic field gradients and pronounced current non-uniformity, the multi-scale approach samples all turns continuously within these boundary regions. For the middle region, an equidistant sampling strategy with an interval of five turns is adopted, ultimately yielding a total of 46 analyzed turns. Regarding mesh discretization, the selected analyzed turns maintain a high-density mesh consistent with the PCR-TA model (40 elements along the width). In contrast, the mesh for the non-analyzed turns is coarsened to 20 elements to optimize computational efficiency.

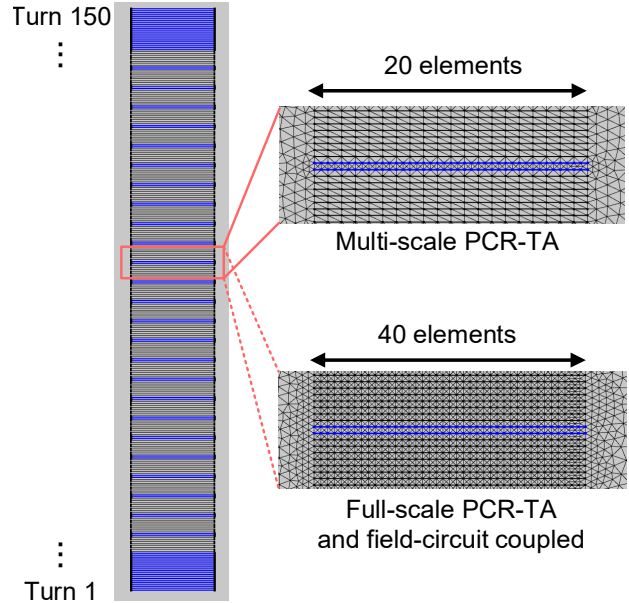

**Figure 8.** Sampling of analyzed turns and mesh discretization in a 150-turn dual-tape PW-NI coil using the multi-scale PCR-TA model. Analyzed turns are shown in blue, while the remaining turns are non-analyzed.

Based on the constructed multi-scale model, simulations were conducted for the 100 A fast-charging and sudden-discharge scenario shown in figure 9. The complete simulation sequence comprises a 5 s ramp-up, a 15 s hold at 100 A, a 0.5 s ramp-down, and a 15 s post-discharge plateau. The time step for the entire simulation was set to 0.01 s. Figure 9 also shows the temporal evolution of the instantaneous magnetization loss computed by the three models over the entire simulation period. The magnetization loss exhibits a local peak at the end of the charging phase ($t$ = 5 s) and a significantly larger peak at the end of the sudden discharge phase ($t$ = 20.5 s). Notably, the curves generated by the multi-scale model closely match those of the PCR-TA and field-circuit coupled models throughout the cycle, providing an initial macroscopic validation of the multi-scale approach.

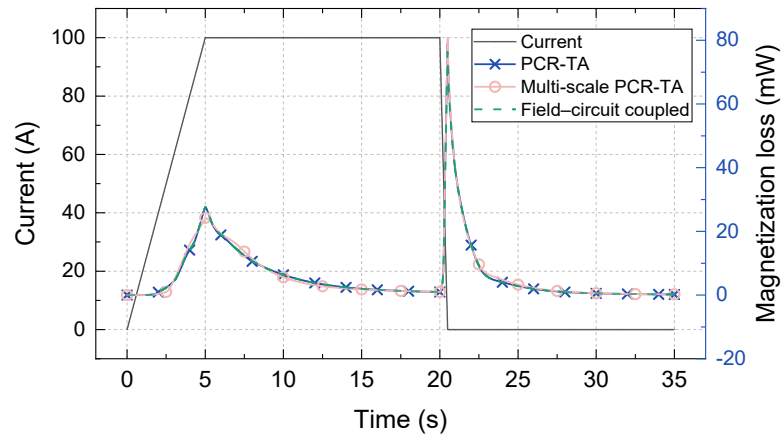

**Figure 9.** Transport current during fast charge/discharge cycles and simulated temporal evolution of losses estimated by the field-circuit coupled model, PCR-TA model, and multi-scale PCR-TA model.

Figure 10 presents the magnetic flux density |$B$| at the steady-state plateau ($t$ = 20 s). It also shows the normalized azimuthal current density $J_\varphi$ distributions under three distinct conditions: the charging peak at $t$ = 5 s, the steady-state plateau at $t$ = 20 s, and the post-discharge residual at $t$ = 25 s. This allows for a qualitative comparison of results from different simulation models. As observed, the multi-scale model successfully reproduces the $J_\varphi$ distributions computed by both the PCR-TA model and the field-circuit coupled model. Owing to this accurate estimation of the $J_\varphi$ distributions, the magnetic field maps obtained from all three models are visually identical.

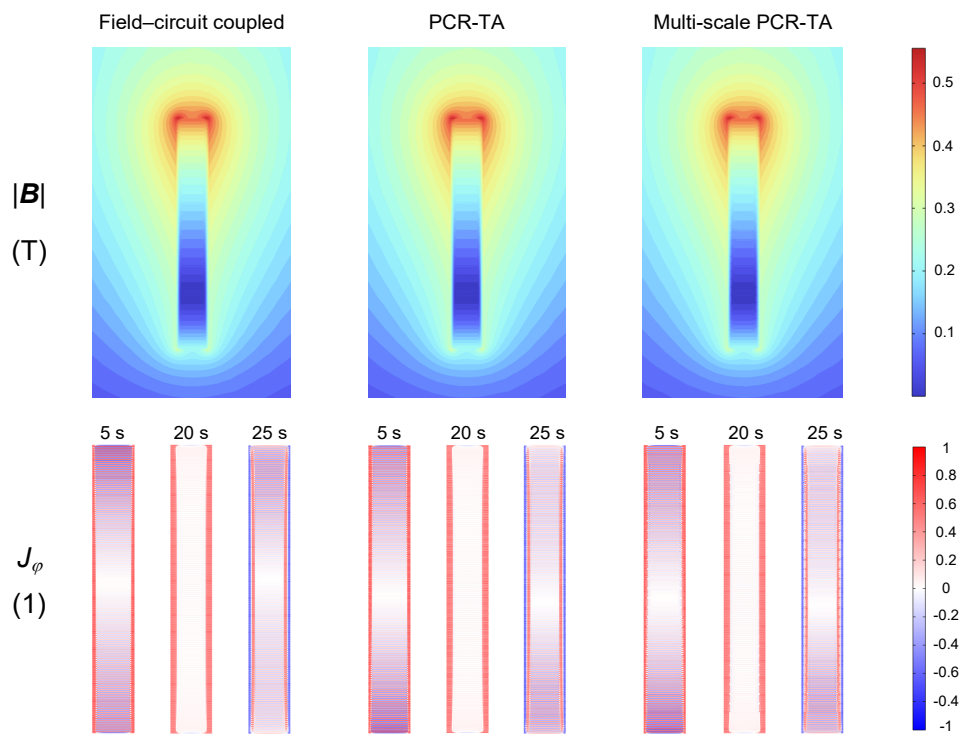

**Figure 10.** Computational results from the field-circuit coupled model, PCR-TA model, and multi-scale PCR-TA model. The top row shows the magnetic flux density at $t$ = 20 s, and the bottom row shows the spatial distribution of azimuthal current density at different time stages.

To quantitatively compare the local computational accuracy, Figure 11 presents the calculated azimuthal and inter-turn radial currents for the 13th turn (a non-analyzed turn) located at the coil edge. In the multi-scale model, the azimuthal current distribution for this turn is obtained via interpolation from the analyzed 10th and 15th turns. It can be seen that the azimuthal and inter-turn radial currents computed by the multi-scale model are highly consistent with the results from the PCR-TA model and the field-circuit coupled model. The multi-scale model accurately reproduces both the pronounced non-uniformity in the azimuthal current evolution and the current sharing within the non-analyzed turns. Compared with the field-circuit coupled model, the maximum deviation occurs at the peak current of Tape B, with an absolute difference of 1.8 A (151.6 A vs. 149.8 A) and a relative error of less than 1.2%. These results demonstrate that the multi-scale approach not only accurately reproduces the macroscopic magnetic field distribution but also effectively captures the detailed evolution of currents in non-analyzed turns.

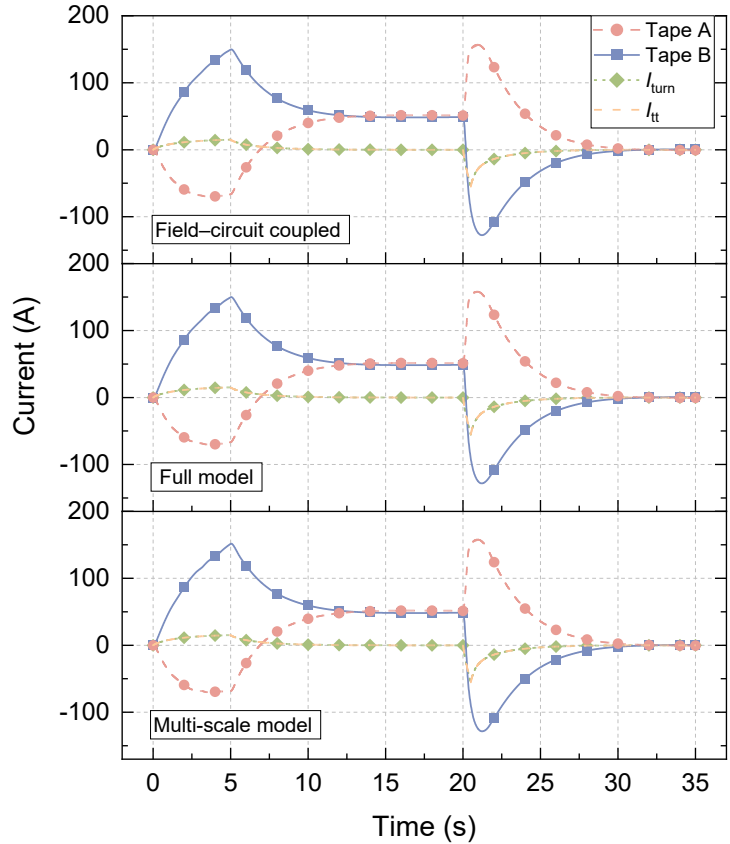

**Figure 11.** Comparison of simulated azimuthal and inter-turn radial currents in the 13th turn (a non-analyzed turn) among the field-circuit coupled model, PCR-TA model, and multi-scale PCR-TA model. The multi-scale model shows high consistency with the PCR-TA model and the field-circuit coupled model.

Table 4 summarizes the integral losses, relative errors, $R^2$ of the current density $J_\varphi$ distributions, and computation times over the 35-s simulation cycle. The losses computed by the PCR-TA model and multi-scale model show good agreement with those from the field-circuit coupled model, with relative errors of 0.42% and 1.16%, respectively. Furthermore, the $R^2$ values for the current densities of both models exceed 0.95, indicating high consistency of the current distributions across the entire spatio-temporal domain.

The most important result is the substantial reduction in computational time achieved by the proposed models. Specifically, the PCR-TA model reduces computational time from 3 h 04 min to 1 h 16 min compared with the field-circuit coupled model, resulting in a speedup factor of approximately 2.4. The multi-scale model further reduces computational time to 32 min, achieving approximately a 5.8-fold speedup compared with the field-circuit coupled model. Furthermore, for engineering applications requiring higher computational efficiency, the multi-scale model can be further simplified by reducing the number of analyzed tapes, particularly in the middle region of the coil, where the current distribution is relatively uniform and has a negligible impact on the total loss. However, excessive reduction of the analyzed turns may introduce cumulative errors, inevitably compromising calculation accuracy.

**Table 4.** Comparison of PCR-TA, multi-scale PCR-TA, and field-circuit coupled models

| Model | Losses (J) | Relative error (%) | $R^2$ of $J_\varphi$ | Computation time |
|---|---|---|---|---|
| Field-circuit coupled model | 0.2158 | — | — | 3 h 04 min |
| PCR-TA model | 0.2164 | 0.42 | 0.9997 | 1 h 16 min |
| Multi-scale PCR-TA model | 0.2180 | 1.16 | 0.9584 | 32 min |

## 4. Extension of the PCR-TA method

### 4.1 Closed-Loop PW-NI Coil

To demonstrate the generality of the proposed model, this section presents its extension to the closed-loop operation simulation of PW-NI coils.

Some groups have simulated closed-loop insulated coils by imposing global integral constraints in the $H$ formulation [47] and applying Neumann boundary conditions in the T–A formulation [48]. In addition, constructing a field-circuit coupled model is another widely adopted strategy, in which an explicit external circuit constrains the electromagnetic field model to achieve a self-consistent simulation of the closed-loop system [49-51]. To simulate the continuous electromagnetic process of the PW-NI coil from power-supply charging to closed-loop operation within the proposed model, a global constraint equation governing the temporal evolution of the total coil current $I_{coil}$ is introduced:

$$(1-\varepsilon(t-t_0))\cdot(I_{coil}-I_{source}(t))+\varepsilon(t-t_0)\cdot(V_{coil}+I_{coil}\cdot R_{cl})=0 \quad (21)$$

where $t_0$ denotes the closed-loop switching time, $\varepsilon(t-t_0)$ is the step function, $I_{source}(t)$ represents the source current, $R_{cl}$ is the short-circuit resistance during closed-loop operation, and $V_{coil}$ denotes the terminal voltage of the coil. Through this global constraint equation, the model enables a self-consistent and seamless transition between the two electromagnetic states. When $t < t_0$, the second term of the equation becomes zero, forcing $I_{coil}$ to follow the $I_{source}(t)$ strictly. When $t \geq t_0$, the first term vanishes, and the equation reduces to the Kirchhoff voltage equation for closed-loop operation. It should be noted that $V_{coil}$ consists of the intrinsic voltage drop of the coil and the joint resistance drops at the inner and outer terminals:

$$V_{coil}=\sum_{i=1}^{N}U_{a,i}+\left(I_{in1}R_{in}(t)+I_{a,N}R_{out}(t)\right) \quad (22)$$

where $R_{in}(t)$ and $R_{out}(t)$ denote the joint resistances at the input and output terminals of Tape A, respectively. Their time dependence accounts for the dynamic evolution of the joint resistances as the system transitions from power-supply charging to closed-loop operation, as expressed in the following equation:

$$R_{in}(t)=\begin{cases} R_{in1}, & t<t_0 \\ R_{in1}+\Delta R, & t\geq t_0 \end{cases} \quad (23)$$

where $R_{in1}$ denotes the static joint resistance during the charging stage defined previously, and $\Delta R$ represents the variation in the joint resistance introduced after switching to the closed-loop operation. Similarly, $R_{out}(t)$ is defined using the same expression.

Based on the proposed model, equation (21) introduces only one additional degree of freedom, enabling continuous simulation of the PW-NI coil over the entire process from power-supply charging to closed-loop operation.

*4.2 Free decay*

To investigate the current decay characteristics of a PW-NI coil under closed-loop conditions, the model is reconfigured by retaining the tape and coil specifications from table 3 while modifying the number of turns to 30. The joint resistances are specified as 300 nΩ during the charging phase and 20 nΩ during closed-loop operation. To clearly demonstrate the current decay, the short-circuit resistance $R_{cl}$ is set to 500 n.

Figure 12 shows the azimuthal currents in the 1st, 10th, and 30th turns of the coil, with the closed-loop switching time $t_0 = 15$ s. Notably, under closed-loop operation, the azimuthal currents of the parallel tapes also exhibit non-uniformity, but the distribution pattern is opposite to that during charging. Specifically, in the innermost 1st turn, Tape B carries the primary current during charging; however, after entering the closed-loop mode, current redistribution occurs, causing the current in Tape A to exceed that in Tape B. The outermost 30th turn exhibits a mirror-symmetric behavior: Tape A dominates during energization, while Tape B carries the larger current during closed-loop operation. Meanwhile, the current transition in the middle 10th turn remains relatively smooth.

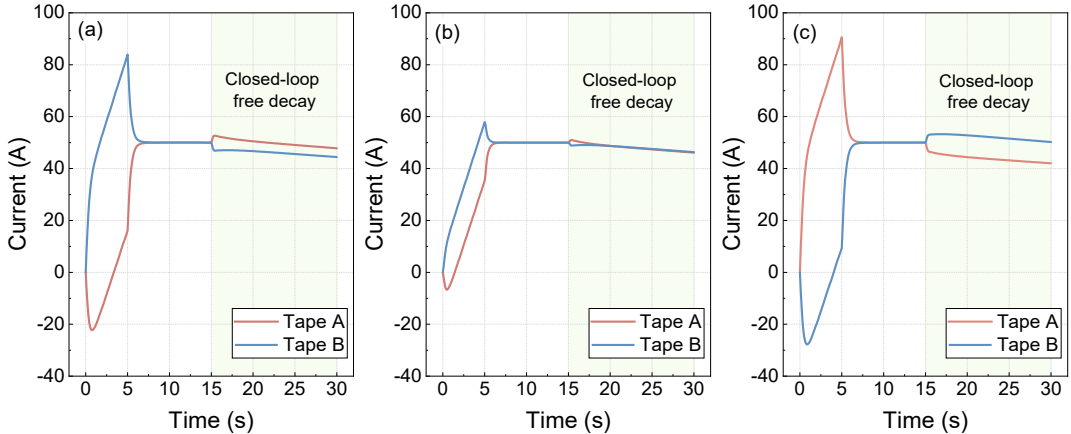

**Figure 12.** Simulated evolution of azimuthal currents in parallel tapes during the transition to closed-loop free decay. (a) 1st turn, (b) 10th turn, and (c) 30th turn. These results illustrate the current redistribution occurring after the closed-loop switching at $t = 15$ s.

The physical mechanism driving this inversion of current distribution stems from the shifting dominance between inductance and contact resistance across the two operational phases. In the inner region, the transient distribution is predominantly governed by contact resistance. During charging, the transport current initially shifts toward Tape B due to radial currents. Upon switching to the closed-loop mode, the abrupt change in electromagnetic boundaries reverses these radial currents, driving the transport current back toward the inner Tape A (as depicted in figure 13(a)). In contrast, current redistribution in the outer turns is primarily dictated by inductance. During charging, Tape A carries more forward current due to its smaller effective inductance.

However, at the transition instant, this smaller inductance forces Tape A to generate a larger reverse coupling current to maintain flux conservation, thereby dropping its net azimuthal current below that of Tape B. To explicitly validate this mechanism, figure 13(b) presents a closed-loop parallel-wound insulated (PW-INS) coil for comparison. By eliminating contact resistance, the current evolution throughout the PW-INS coil is predominantly governed by inductance, closely resembling that of the outer turns in the PW-NI coil. This confirms that the current distribution reversal observed in the inner turns of the PW-NI coil is primarily driven by contact resistance, whereas the outer turns are dominated by inductance.

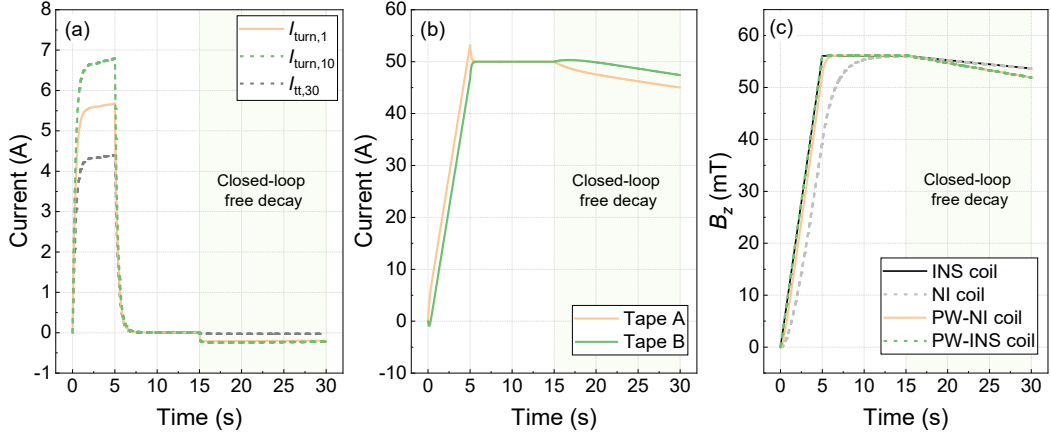

**Figure 13.** Time evolution of simulated currents and central magnetic field for different coil configurations (a) inter-turn radial currents in the 1st, 10th, and 30th turns of the PW-NI coil, (b) azimuthal currents in the PW-INS coil, and (c) comparison of central magnetic field evolution for different coil configurations.

Figure 13(c) compares the evolution of the central magnetic field for the dual-tape PW-NI, dual-tape PW-INS, single-tape NI, and single-tape INS coils, all of which share identical geometric dimensions and the same number of winding layers. During charging, the central magnetic fields of the INS and PW-INS coils closely follow the external excitation current due to the inter-layer insulation. The PW-NI coil exhibits only a slight delay, as its parallel-wound structure significantly reduces the effective inductance. In contrast, the single-tape NI coil shows a substantial charging delay. Compared to the parallel-wound configurations, the single-tape INS and NI coils demonstrate a more gradual decay after closed-loop activation. This difference stems from the lower equivalent inductance of the parallel-wound coils at equivalent ampere-turns, which shortens their closed-loop time constants. Notably, although the PW-NI coil undergoes complex internal current redistribution at the instant of closed-loop activation, its macroscopic magnetic field decay closely matches that of the PW-INS coil. This indicates that the localized current distribution reversals do not compromise the overall axial magnetic field.

*4.3 Current decay under a radial AC field*

To investigate the current decay characteristics of the closed-loop PW-NI coil under a radial AC magnetic field, a pair of background coils connected in anti-series was positioned directly above and below the bifilar coil, as shown in figure 14. This configuration enables the axial magnetic fluxes generated by the two background coils to cancel each other, thereby producing an alternating magnetic field predominantly distributed radially within the superconducting coil. Except for the AC excitation, all other system parameters and the charging procedure were kept identical to those under the free-decay conditions described in Section 4.1. After the completion of the charging process, the closed-loop free-decay phase was initiated at $t = 15$ s, followed by the application of AC excitation to the background coils at $t = 30$ s. This excitation had an amplitude of 50 A and a frequency of 1 Hz, producing a peak radial magnetic field of 16.7 mT at the center of the PW-NI coil.

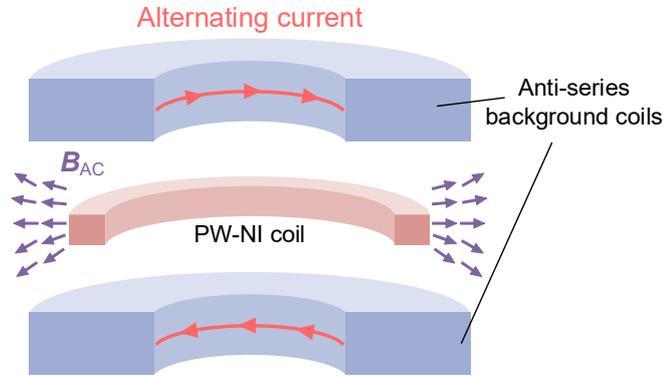

**Figure 14.** Schematic of a closed-loop PW-NI coil subjected to an external radial AC magnetic field.

Figures 15(a) and 15(b) illustrate the dynamic evolution of currents in the parallel tapes and the total azimuthal current in the 1st and 30th turns of the PW-NI coil, respectively. During the first cycle after applying the AC magnetic field, the total azimuthal current exhibits a pronounced and rapid decay. This behavior results from asymmetric magnetic flux penetration at the initial application of the field, which introduces additional resistance into the closed loop [52]. After the first-cycle impact subsides ($t > 31$ s), the current enters a smooth decay phase. The persistent AC field continuously induces local shielding currents within the tapes, producing dynamic resistance that further accelerates the closed-loop current decay.

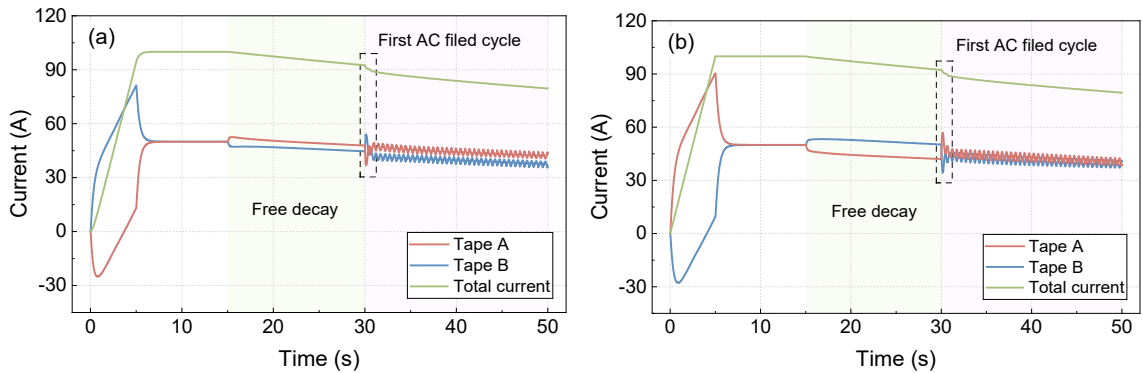

**Figure 15.** Transient evolution of azimuthal currents in parallel tapes and total current under an external radial AC magnetic field disturbance: (a) Currents in the 1st turn, (b) Currents in the 30th turn.

As shown in Figure 15, the AC magnetic field induces significant current redistribution between the parallel tapes. During the initial AC cycle (30 s < $t$ < 31 s), the 1st turn exhibits a negative current spike in Tape A and a positive spike in Tape B, while the 30th turn displays the opposite trend. This transient response is driven by local dynamic resistance arising from alternating flux penetration. Specifically, figures 16(a) and 16(b) reveal that tapes at the electromagnetic boundaries (i.e., Tape A in the 1st turn and Tape B in the 30th turn) exhibit notably higher dynamic resistance peaks during this first cycle. This phenomenon results from the strongest initial flux penetration occurring at these boundary regions, which directly forces the current into the other parallel tape. After the first cycle, radial positional differences between the tapes cause the AC field to generate asymmetric dynamic resistance continuously. This sustains inter-tape current redistribution, accompanied by fluctuations at twice the applied AC frequency and a concurrent decay of the closed-loop current.

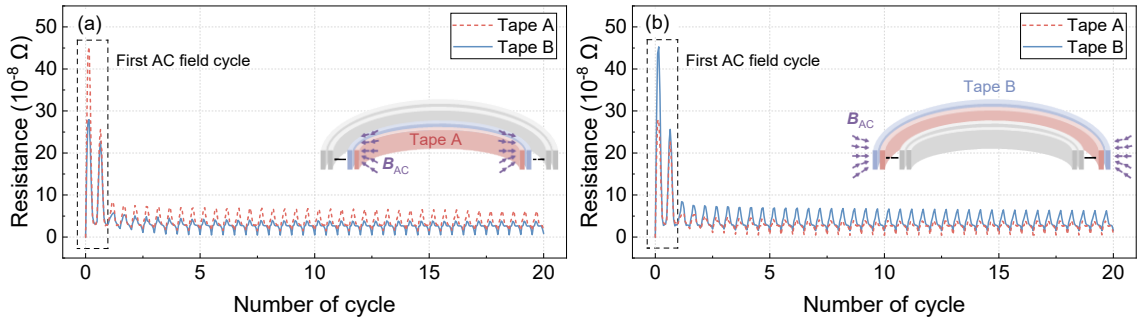

**Figure 16.** Asymmetric evolution of transient dynamic resistance in parallel tapes under an external radial AC magnetic field. (a) Dynamic resistance in the 1st turn, where Tape A (at the inner boundary) exhibits a higher peak. (b) Dynamic resistance in the 30th turn, where Tape B (at the outer boundary) exhibits a higher peak.

## 5. Conclusion

In this work, an efficient electromagnetic simulation method for PW-NI coils is developed based on the $T$-$A$ formulation by introducing the potential-chain recursion among parallel tapes. This method directly embeds the current sharing and radial shunting behaviors among parallel tapes into the finite element framework, thereby eliminating the need for equivalent circuit models or cross-module data interactions. Furthermore, this framework can be extended to simulate pancake coils with various winding configurations. Building upon this foundation, a multi-scale approach is further developed for the electromagnetic simulation of large-scale PW-NI coils.

The proposed method is validated against experimental measurements of dual- and three-tape PW-NI coils, as well as simulation results from a field-circuit coupled model based on the $T$-$A$ formulation. Comparative analyses show that the PCR-TA method achieves an approximately 2.4-fold computational speedup over the field-circuit coupled approach, while the multi-scale approach further increases the speedup to roughly 5.8-fold, without compromising the accuracy of local current density distributions or associated losses.

The proposed model is extended to enable continuous simulation of PW-NI coils from power-supply charging to closed-loop operation. Simulation results show that, upon closed-loop formation, the current redistributes between the parallel tapes, leading to spatial non-uniformity of the azimuthal current. Furthermore, under an external radial AC magnetic field, the closed-loop current initially decays rapidly due to asymmetric magnetic flux penetration. Subsequently, the AC field continuously induces asymmetric transient dynamic resistance in the parallel tapes, driving further current redistribution and generating current fluctuations at twice the applied AC frequency. Future work will focus on extending the proposed method to 3D full-scale modeling and multi-physics coupling, establishing a comprehensive and efficient numerical framework for large-scale HTS magnet systems.

**Data and materials availability**

All data that support the findings of this study are available upon reasonable request from the authors.

**Acknowledgments**

This work was supported by the National Key Research and Development Program of China (Grant No. 2024YFF0508002-05). The authors would like to acknowledge Sizuo Chen for his assistance with the experimental setup.

## References

[1] Hahn S, Park D K, Bascunan J, Iwasa Y 2011 *IEEE Trans. Appl. Supercond.* **21** 2093492

[2] Wang Q, Liu J, Zheng J, Qin J, Ma Y, Xu Q, Wang D, Chen W, Qu T, Zhang X *et al.* 2022 *Supercond. Sci. Technol.* **35** 023001

[3] Dong F, Park D, Sadde P, Bascunan J, Iwasa Y 2025 *Supercond. Sci. Technol.* **38** 035016

[4] Liu J, Wang Q, Qin L, Zhou B, Wang K, Wang Y, Wang L, Zhang Z, Dai Y, Liu H *et al.* 2020 *Supercond. Sci. Technol.* **33** 3LT01

[5] Li Y 2025 *Supercond. Sci. Technol.* **38** 063001

[6] Gao C, Chen P H, Alaniva N, Ellison J H J, Frei M, Björgvinsdóttir S, Saliba E P, Hu Y, Pagonakis I, Däpp A *et al.* 2026 *Sci. Adv.* **12** eadz5826

[7] Parkinson B J 2017 *Supercond. Sci. Technol.* **30** 014009

[8] Bhattarai K R, Kim K, Kim S, Lee S G and Hahn S 2017 *IEEE Trans. Appl. Supercond.* **27** 4603505

[9] Noguchi S 2019 *IEEE Trans. Appl. Supercond.* **29** 4602607

[10] Choi Y H, Kim S-G, Jeong S-H, Kim J H, Kim H M, Lee H 2017 *IEEE Trans. Appl. Supercond.* **27** 4601206

[11] Wang Y, Song H, Xu D, Li Z Y, Jin Z and Hong Z 2015 *Supercond. Sci. Technol.* **28** 045017

[12] Yoshihara Y, Hamanaka M, Tsuyoshi K, Kitamura M, Nemoto U, Noguchi S and Ishiyama A 2021 *IEEE Trans. Appl. Supercond.* **31** 4602005

[13] Xue W, Huang Z, Xu X, Shen B and Jin Z 2021 *IEEE Trans. Appl. Supercond.* **31** 4902105

[14] Kim K, Bhattarai K, Kim K L, Bai H, Dixon I R, Painter T A, Bong U, Larbalestier D C and Hahn S 2020 *IEEE Trans. Appl. Supercond.* **30** 4602205

[15] Geng J and Zhang M 2019 *Supercond. Sci. Technol.* **32** 084002

[16] Miyamoto Y, Ishiyama A, Watanabe T and Nagaya S 2022 *IEEE Trans. Appl. Supercond.* **32** 5700605

[17] Ogasawara T and Ishiyama A 2024 *IEEE Trans. Appl. Supercond.* **34** 5700905

[18] Sato H, Mato T, Obana T and Noguchi S 2025 *IEEE Trans. Appl. Supercond.* **35** 4902605

[19] Chang Z, Tang Y, Li D, Liu D, Yong H, Gao Z, Sun S 2025 *Appl. Therm. Eng.* **280** 128387

[20] Liu J, Wang Q, Zhou B, Zhang Z, Xu C, Chen S, Sun W, Chen Y, Ji X, Zeng H *et al.* 2026 *Superconductivity* **17** 100240

[21] Hartwig Z S, Vieira R F, Dunn D, Golfinopoulos T, LaBombard B, Lammi C J, Michael P C, Agabian S, Arsenault D, Barnett R, et al 2024 *IEEE Trans. Appl. Supercond.* **34** 0600316

[22] Golfinopoulos T, Michael P C, Ihloff E, Zhukovsky A, Nash D, Fry V, Muncks J P, Barnett R, Bartoszek L, Beck W, et al 2024 *IEEE Trans. Appl. Supercond.* **34** 0600416

[23] Gryaznevich M 2022 *Nucl. Fusion* **62** 042008

[24] Kobayashi H, Nakada Y, Saotome H, Miyagi D, Nagasaki Y and Tsuda M 2023 *IEEE Trans. Appl. Supercond.* **33** 4601606

[25] Lee J T, Kim G, Park H, Park J, Im C, Jang W, Kim J, Hahn S 2026 *IEEE Trans. Appl. Supercond.* **36** 4604505

[26] Brambilla R, Grilli F, Martini L 2007 *Supercond. Sci. Technol.* **20** 16–24

[27] Mataira R C, Ainslie M D, Badcock R A, Bumby C W 2020 *Supercond. Sci. Technol.* **33** 08LT01

[28] Koshy B G, Ainslie M, Sun Y, Mallett B P P, Jiang Z 2025 *Superconductivity* **16** 100216

[29] Zhou B, Wang L, Chen Y, Wang Q, Wang K, Zhang Z, Liu J 2024 *Science China Technol. Sci.* **67** 2154–2166

[30] Liang F, Venuturumilli S, Zhang H, Zhang M, Kvitkovic J, Pamidi S, Wang Y, Yuan W 2017 *J. Appl. Phys.* **122** 043903

[31] Berrosp E, Zermeño V, Trillaud F, Grilli F 2019 *Supercond. Sci. Technol.* **32** 065003

[32] Ji X, Chen Y, Zhou B, Wang Q, Liu J 2025 *IEEE Trans. Appl. Supercond.* **35** 3583430

[33] Fu Y, Wang Y, Peng W, Zhao Y, Ma G and Jin Z 2023 *Supercond. Sci. Technol.* **36** 115031

[34] Fu Y, Ma G, Dong F, Wang Y 2024 *Superconductivity* **12** 100140

[35] Liu Y, Song P, Xiao M, Shao L, Xu Z, Korte C and Qu T 2025 *Supercond. Sci. Technol.* **38** 125005

[36] Chen Y, Wang Q, Wang K, Zhou B, Zeng H, Liu S, Ji X, Wang L, Liu J 2025 *Superconductivity* **16** 100217

[37] Chen Y, Wang Q, Zhou B, Liu S, Dong F, Wang K, Zeng H, Ji X, Wang L, Liu J *et al.* 2026 *Supercond. Sci. Technol.* **39** 025005

[38] Liu X, Fu Y, Peng W, Zhao Y, Wang Y 2025 *IEEE Trans. Appl. Supercond.* **35** 4603107

[39] Zeng H, Wang Q, Chen Y, Wang K, Zhou B, Ji X 2026 *IEEE Trans. Appl. Supercond.* **36** 4604012

[40] Huber F, Song W, Zhang M and Grilli F 2022 *Supercond. Sci. Technol.* **35** 043003

[41] Kim Y B, Hempstead C F and Strnad A R 1963 *Phys. Rev.* **129** 528–535

[42] Wang Y, Chan W K and Schwartz J 2016 *Supercond. Sci. Technol.* **29** 045007

[43] Su B, Zhou W, Liu W, Jia R, Wang N, Chen T, Liang R 2025 *Physica C* **637** 1354785

[44] Chen Y, Wang Q, Wang K, Zhou B, Liu S, Ji X, Wang L, Liu J 2024 *Supercond. Sci. Technol.* **37** 095012

[45] Wang X, Hahn S, Kim Y, Bascuñán J, Voccio J, Lee H and Iwasa Y 2013 *Supercond. Sci. Technol.* **26** 035012

[46] Zhou P, Dos Santos G, Ghabeli A, Grilli F and Ma G 2022 *Supercond. Sci. Technol.* **35** 115005

[47] Zhong Z, Wu W, Wang L, Li X-F, Li Z, Hong Z and Jin Z 2021 *J. Supercond. Nov. Magn.* **34** 2809

[48] Wang S, Yong H, Zhou Y 2022 *Supercond. Sci. Technol.* **35** 065013

[49] Zhou P, Zhang S, Wang R, Li S, Grilli F and Ma G 2024 *Supercond. Sci. Technol.* **37** 065001

[50] Zhu L, Wang Y, Guo Y, Liu W and Hu C 2023 *Energy* **277** 127560

[51] Lu L, Wu W, Gao Y, Pan C, Yu X, Zhang C, Jin Z 2022 *Supercond. Sci. Technol.* **35** 095001

[52] Zhong Z, Wu W and Jin Z 2021 *Supercond. Sci. Technol.* **34** 08LT01